\documentclass[pra,twocolumn,showpacs,superscriptaddress]{revtex4-1}
\usepackage{graphicx}
\usepackage{amssymb}
\usepackage{amsfonts}
\usepackage{amsmath}
\usepackage{lmodern}
\usepackage{mathrsfs}
\usepackage[dvipdfm,colorlinks=true, citecolor=blue, urlcolor=blue ]{hyperref}
\begin{document}

\title{Generation of stable entanglement between two cavity mirrors by squeezed-reservoir engineering}
\author{Chun-Jie Yang}
\affiliation{Center for Interdisciplinary Studies $\&$ Key Laboratory for Magnetism and Magnetic Materials of the MoE, Lanzhou University, Lanzhou 730000, China}
\author{Jun-Hong An}\email{anjhong@lzu.edu.cn}
\affiliation{Center for Interdisciplinary Studies $\&$ Key Laboratory for Magnetism and Magnetic Materials of the MoE, Lanzhou University, Lanzhou 730000, China}
\author{Wanli Yang}
\affiliation{State Key Laboratory of Magnetic Resonance and Atomic and Molecular Physics, Wuhan Institute of Physics and Mathematics, Chinese Academy of
Sciences, Wuhan 430071, China}
\author{Yong Li}
\affiliation{Beijing Computational Science Research Center, Beijing 100084, China}

\begin{abstract}
The generation of quantum entanglement of macroscopic or mesoscopic bodies in mechanical motion is generally bounded by the thermal fluctuation exerted by their environments. Here we propose a scheme to establish stationary entanglement between two mechanically oscillating mirrors of a cavity. It is revealed that, by applying a broadband squeezed laser acting as a squeezed-vacuum reservoir to the cavity, a stable entanglement between the mechanical mirrors can be generated. Using the adiabatic elimination and master equation methods, we analytically find that the generated entanglement is essentially determined by the squeezing of the relative momentum of the mechanical mirrors, which is transferred from the squeezed reservoir through the cavity. Numerical verification indicates that our scheme is within the present experimental state of the art of optomechanics.
\end{abstract}

\maketitle

\section{Introduction}

Quantum entanglement, as a cornerstone in understanding many phenomena in quantum world \cite{Osterloh2002,Zurek2003,Lambert2004,Dur2005,Popescu2006}, serves as a necessary resource in various practical applications of quantum information processing, such as quantum algorithms \cite{Childs2010}, quantum teleportation \cite{Jin2010}, and quantum crytography \cite{Gisin2010}. In the past decade, tremendous efforts have been devoted to generate entanglement in microscopic systems \cite{Bayer2001,Julsgaard2001,Zhao2004,Bowen2004,Blinov2004,Leibfried2005,Haffner2005,Jost2009,Yao2012,Cramer2013}. Recently, the generation of entanglement in macroscopic and mesoscopic objects and the study of quantum mechanical features in these scales have attracted much attentions \cite{Vedral2008,Amico2008}.

An optomechanical system supplies an ideal platform to explore quantum features of macroscopic or mesoscopic objects in mechanical motion \cite{Aspelmeyer2014}.  Advances in this field raise a fundamental question: Whether mechanical systems in macroscopic scale exhibit quantum behavior? People desire to see under what conditions it is feasible to generate nonclassical entangled states in macromechanical oscillators. It was found that, similar to the microscopic system case, entanglement between remote mirrors can also be generated via optical measurement based on the entanglement swapping idea \cite{Mancini2002,Pirandola2006,Vacanti2008,Borkje2011} and via the coherent interactions between the fields in two adjacent cavities \cite{Joshi2012,Chen2014,Liao2014}. However, the emergence of quantum effects in macroscopic objects is generally believed to be bounded by the thermal fluctuation. Therefore, schemes resorting to an efficient precooling to the thermal noise have been proposed to establish a stable entanglement between the mirrors of a cavity \cite{Vitali2003,Genes2008} and between the two dielectric membranes suspended inside a cavity \cite{Hartmann2008}. Further studies showed that the reservoir engineering technique supplies a nice idea to entangle the mechanical systems without resorting to precooling. For example, it is found that a stable entanglement between the mechanical oscillators in separated cavities \cite{Tan2013} and between the cavity mirror and atomic ensemble \cite{Hammerer2009} can be generated based on the cascade input-output process. Other schemes based on engineering the squeezing characters of the reservoirs have also been proposed to entangle the mechanical mirrors in separated cavities \cite{Zhang2003,Sete2014}, in ring-cavity \cite{Agarwal2009}, and in double-cavity \cite{Pinard2005} setups. A method using multiple-tone coherent driving to the cavity has been used to entangle two mechanical mirrors for a single cavity \cite{Woolley2014, Li2015}. %However, in these works, only qualitative entanglement analysis based on Duan inequality \cite{Duan} was present, which is difficult for evaluating the physical condition to get an optimal entanglement.

Inspired by these obvious benefits, i.e. the robustness to thermal noise and no precooling, of the reservoir engineering in quantum optomechanical control \cite{Tan2013,Hammerer2009,Zhang2003,Sete2014,Agarwal2009,Pinard2005}, we in this work propose a scheme to stably entangle the two mechanical mirrors of a single Fabry-Perot cavity via the squeezed-reservoir engineering. Going from the master equation of the whole system and adiabatically eliminating the degree of freedom of the cavity field, we derive a reduced master equation satisfied by the mechanical oscillators. Our analytic study on the mechanical entanglement quantitatively characterized by logarithmic negativity reveals that the entanglement generated comes from the squeezing of the relative momentum of the two mechanical oscillators. A temperature dependent entanglement criterion is obtained, which shows the thermal fluctuation tolerance of the generated entanglement. The physical condition for achieving the maximal entanglement is explicitly obtained from our analysis.

Our paper is organized as follows. In Sec. \ref{sys}, we show the model and derive the reduced master equation by adiabatically eliminating the cavity mode. In Sec. \ref{ent}, the entanglement generation between the mechanical oscillators is explicitly studied. The applicability of the adiabatic elimination is also verified. In Sec. \ref{con}, a summary is given.

\section{System and Adiabatic Elimination}\label{sys}

\begin{figure}[tbp]
\centering
\includegraphics[width=0.6\columnwidth]{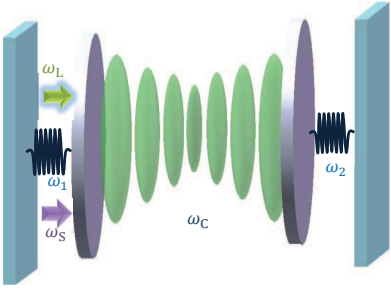}
\caption{(Color online) Schematic diagram of a Fabry-Perot cavity with two mechanically oscillating mirrors in frequencies $\omega_1$ and $\omega_2$. A coherent laser with frequency $\omega _{\text{L}}$ and a broadband squeezed laser with squeezing degree $r$ around the central frequency $\omega_\text{S}$ are injected into the cavity to implement the stable entanglement generation between the mechanical oscillators. }\label{Fig10}
\end{figure}
We consider a laser in frequency $\omega _{\text{L}}$ driven cavity with two oscillating mirrors (see Fig. \ref{Fig10}) \cite{Woolley2014, Li2015}. The ideal situation without thermal noise of this system has been studied in Refs. \cite{Huang2012}. The Hamiltonian of the total system is ($\hbar =1$)
\begin{equation}
\hat{H}=\Delta _{\text{c}}\hat{c}^{\dag }\hat{c}+\sum_{j=1,2}[\omega _{j}\hat{a}_{j}^{\dag }\hat{a}_{j}+\eta _{j}(\hat{a}_{j}^{\dag }+\hat{a}_{j})\hat{c}^{\dag }\hat{c}]+\Omega (\hat{c}^{\dag }+\hat{c}),  \label{Hamini}
\end{equation}
where $\hat{c}$ and $\hat{a}_{j}$, respectively, denote the annihilation operators of the cavity field and the two mechanical oscillators formed by the mirrors, $\eta _{j}$ denote the coupling strengths between the cavity field and the oscillators due to the radiation pressure, $\Delta _{\text{c}}=\omega _{\text{c}}-\omega _{\text{L}}$ is the detuning of the laser frequency $\omega_\text{L}$ to the cavity-field frequency $\omega_\text{c}$, and $\Omega =2\sqrt{P\kappa/ \omega _\text{L}}$ is the driving amplitude with the input laser power $P$ and the cavity damping rate $\kappa $. The mechanical oscillators interact with two independent reservoirs at same temperature $T$. The cavity is further driven by a broadband squeezed laser with squeezing degree $r$ around the central frequency $\omega _{\text{S}}$, which acts as a squeezed-vacuum reservoir to the cavity field. Then the total system is governed by the Born-Markovian master equation \cite{Scully1997}
\begin{equation}
\dot{W}(t)=-i[\hat{H},W(t)]+\mathcal{\hat{L}}_{\text{m}}W(t)+\mathcal{\hat{L}}_{\text{c}}W(t),  \label{mastini}
\end{equation}
where $W(t)$ is density matrix of the total system, $\mathcal{\hat{L}}_{\text{m}}\cdot =\sum_{j=1,2}\gamma _{j}[(\bar{n}_{j}+1)\check{{\mathcal{D}}}_{\hat{a}_{j},\hat{a}_{j}^{\dag }}\cdot +\bar{n}_{j}\check{{\mathcal{D}}}_{\hat{a}_{j}^{\dag },\hat{a}_{j}}\cdot ]$ with $\bar{n}_{j}=1/[\exp (\omega_{j}/k_{B}T)-1]$ and $\check{\mathcal{D}}_{\hat{o},\hat{p}}\cdot =2\hat{o}\cdot \hat{p}-\hat{p}\hat{o}\cdot -\cdot \hat{p}\hat{o}$ represents the dissipators of the two mirrors caused by their two independent finite-temperature reservoirs, $\mathcal{\hat{L}}_{\text{c}}\cdot =\kappa\lbrack (N+1)\check{{\mathcal{D}}}_{\hat{c},\hat{c}^{\dag }}\cdot +N\check{{\mathcal{D}}}_{\hat{c}^{\dag },\hat{c}}\cdot -(Me^{i2\Delta _{\text{s}}t}\check{{\mathcal{D}}}_{\hat{c},\hat{c}}\cdot +$h.c.$)]$, with $\Delta_\text{s}=\omega_\text{S}-\omega_\text{L}$, $M=\cosh r\sinh r$, and $N=\sinh ^{2}r$ represents the dissipator of the cavity field caused by the squeezed-vacuum reservoir. $\gamma _{j}$ and $\kappa$ are the damping rates of the cavity and the oscillators, respectively. From Eq. (\ref{mastini}), the steady state value of the cavity field and the oscillators can be obtained as $\langle \hat{c}\rangle _{\text{ss}}=\Omega /(i\kappa -\Delta _{\text{c}})\equiv \alpha $ and $\langle \hat{a}_{j}\rangle _{\text{ss}}=-\eta _{j}|\alpha |^{2}/(\omega_{j}-i\gamma _{j})$. Then Eq. (\ref{Hamini}) can be linearized into $\hat{H}=\hat{H}_{0}+\hat{H}_{\text{I}}$, where $\hat{H}_{0}=\sum_{j=1,2}\omega _{j}\hat{a}_{j}^{\dag }\hat{a}_{j}+\Delta _{\text{c}}\hat{c}^{\dag }\hat{c}$ and $\hat{H}_{\text{I}}=\sum_{j=1,2}\eta _{j}(\hat{a}_{j}^{\dag }+\hat{a}_{j})(\alpha \hat{c}^{\dag }+\alpha ^{\ast }\hat{c})$.

In the large damping limit $\kappa\gg\gamma_j$, we can adiabatically eliminate the degree of freedom of the cavity field and get a reduced master equation satisfied by the two mechanical oscillator (see Appendix \ref{derimse})
\begin{eqnarray}
&&\dot{\rho}(t)=-i[\sum_{j=1,2}\omega _{j}\hat{a}_{j}^{\dag }\hat{a}_{j},\rho (t)]+\mathcal{\hat{L}}_{\text{m}}\rho (t)+\sum_{j,k=1,2}\eta_{j}\eta _{k}  \notag \\
&&\times \lbrack (\xi _{j}^{+\ast }+\xi _{k}^{+})\hat{a}_{j}\rho (t)\hat{a}_{k}^{\dag }-\xi _{j}^{+\ast }\hat{a}_{k}^{\dag }\hat{a}_{j}\rho (t)-\xi_{k}^{+}\rho (t)\hat{a}_{k}^{\dag }\hat{a}_{j}  \notag \\
&&+(\xi _{j}^{-\ast }+\xi _{k}^{-})\hat{a}_{j}^{\dag }\rho (t)\hat{a}_{k}-\xi _{j}^{-\ast }\hat{a}_{k}\hat{a}_{j}^{\dag }\rho (t)-\xi_{k}^{-}\rho (t)\hat{a}_{k}\hat{a}_{j}^{\dag }]  \notag \\
&&+(\xi _{j}^{+\ast }+\xi _{k}^{-})\hat{a}_{j}\rho (t)\hat{a}_{k}-\xi_{j}^{+\ast }\hat{a}_{k}\hat{a}_{j}\rho (t)-\xi _{k}^{-}\rho (t)\hat{a}_{k}
\hat{a}_{j}  \notag \\
&&+(\xi _{j}^{-\ast }+\xi _{k}^{+})\hat{a}_{j}^{\dag }\rho (t)\hat{a}_{k}^{\dag }-\xi _{j}^{-\ast }\hat{a}_{k}^{\dag }\hat{a}_{j}^{\dag }\rho
(t)-\xi _{k}^{+}\rho (t)\hat{a}_{k}^{\dag }\hat{a}_{j}^{\dag },\label{remast}
\end{eqnarray}%
where $\rho(t)=\text{Tr}_\text{c}[W(t)]$, $\xi _{k}^{\pm }=\frac{\digamma }{\kappa +i(\Delta\pm\omega _{k})}+\frac{|\alpha |^{2}+\digamma ^{\ast }}{\kappa -i(\Delta\mp \omega _{k})}$ with $\Delta\equiv\Delta_\text{s}=\Delta_\text{c}$ and $\digamma =N|\alpha |^{2}+M\alpha^{2}e^{2i\Delta t}$. Keeping a traceless structure, Eq. (\ref{remast}) preserves the positivity of the reduced density matrix $\rho(t)$.

The newly emergent third term in the right-hand side of Eq. (\ref{remast}) incorporates all the dynamical effects of the cavity field on the two mechanical oscillators. It is interesting to see that the cavity field, as a common contact ``environment", can not only induce individual dissipation (with $j=k$) to each mirror, but also induce incoherent interactions (with $j\neq k$) between the two mirrors by the exchange of virtual phonons. Furthermore, besides the thermal dissipation [the second and third lines of Eq. (\ref{remast})], the squeezing-like dissipation [the fourth and fifth lines of Eq. (\ref{remast})] can also be triggered. It is understandable based on the existence of the counter-rotating terms in $\hat{H}_\text{I}$. In the special case of $r=0$ and single oscillating mirror, after dropping the fast rotating squeezing-dissipation terms \cite{Wilson2008}, Eq. \eqref{remast} reduces exactly to the similar form as the one in Ref. \cite{Kippenberg2007}. In this case, the rates of cooling and heating denoted by the second and third lines of Eq. \eqref{remast} reduce to $(\xi_j^{+*}+\xi_j^+)|_{r=0}\propto [\kappa^2+(\Delta-\omega_j)^2]^{-1}$ and $(\xi_j^{-*}+\xi_j^-)|_{r=0}\propto [\kappa^2+(\Delta+\omega_j)^2]^{-1}$, respectively. It means that an efficient cooling is realizable when the cooling rate is larger than the heating one by choosing red-detuning driving field (i.e. $\Delta>0$ ) \cite{Kippenberg2007}. When $r\neq0$, the incoherent interactions and the squeezing effect induced by the cavity field can cause the oscillators in squeezed state (see Appendix \ref{app-dynamics}), which is crucial for generating stable entanglement between the two thermally oscillating mirrors as shown in the following.

Considering explicitly the scheme configuration in Fig. \ref{Fig10}, we have the physical condition that the cavity field interacts with the two mirrors in a $\pi$-phase difference $\eta _{1}=-\eta_{2}\equiv\eta _{0}$. Further assuming the two mirrors are identical, we get $\omega_1=\omega_2\equiv\omega_0$, $\gamma _{1}=\gamma_{2}\equiv\gamma _{0}$, and $\bar{n}_{1}=\bar{n}_{2}\equiv\bar{n}_{0}$. Then Eq. (\ref{remast}) reduces to
\begin{eqnarray}
&&\dot{\rho}(t) =-i[\hat{H}_\text{eff},\rho (t)]\nonumber\\
&&+\sum_{j=\pm}\gamma_0[(\bar{n}_0+1)\check{\mathcal{D}}_{\hat{a}_j,\hat{a}^\dag_j}\rho(t)+\bar{n}_0\check{\mathcal{D}}_{\hat{a}^\dag_j,\hat{a}_j}\rho(t)]\notag \\
&&+\eta _{0}^{2}[(\xi _{0}^{+\ast }+\xi _{0}^{+})\check{\mathcal{D}}_{\hat{a}_-,\hat{a}^\dag_-}\rho(t) +(\xi _{0}^{-\ast }+\xi _{0}^{-})\check{\mathcal{D}}_{\hat{a}^\dag_-,\hat{a}_-}\rho(t) \notag\\
&&+(\xi _{0}^{+\ast }+\xi _{0}^{-})\check{\mathcal{D}}_{\hat{a}_-,\hat{a}_-}\rho(t)+(\xi _{0}^{-\ast }+\xi _{0}^{+})\check{\mathcal{D}}_{\hat{a}^\dag_-,\hat{a}^\dag_-}\rho(t)],\label{speredms}
\end{eqnarray}
where $\hat{a}_{\pm }=(\hat{a}_{1}\pm\hat{a}_{2})/\sqrt{2}$ and $\hat{H}_{eff}=\sum_{j=\pm }\omega _{j}\hat{a}_{j}^{\dag }\hat{a}_{j}+[i\eta _{0}^{2}(\xi _{0}^{-}-\xi _{0}^{+\ast })\hat{a}_{-}^2+\text{h.c.}]$ with $\omega _{+}=\omega _{0}$ and $\omega _{-}=\omega _{0}-2\eta _{0}^{2}$Im$[\xi _{0}^{+}+\xi _{0}^{-}]$. It can be seen that the relative motion degree of freedom of the two mirrors is decoupled to the center-of-mass one and only the former feels the existence of the cavity environment. The presence of the squeezing Lindblad terms in the last terms of Eq. (\ref{speredms}) inspires us that a stable squeezing property could be induced to the relative motion degree of freedom by the cavity environment, via which a stable entanglement is hopefully attainable.

%We here will show that the incoherent interactions and the squeezing effect induced by the common cavity field can be used to generate stable quantum entanglement between the two thermally oscillating mechanical mirrors. It is noted that the coefficients of Eq. (\ref{remast}) is time-dependent, which comes from the squeezed driving field.

\section{Entanglement between mechanical oscillators}\label{ent}
\subsection{Analytical criterion for the steady-state entanglement}
%\subsection{Quantum entanglement and quantum fluctuations}
The entanglement of our system can be measured by the logarithmic negativity \cite{Vidal2002}, which is quantified on the covariance matrix of $\mathbf{\hat{{X}}}\equiv(\hat{{x}}_{1},{\hat{p}}_{1},\hat{{x}}_{2},\hat{{p}}_{2})$,
\begin{equation}
{V}_{ij}=\langle \Delta \hat{{X}}_{i}\Delta \hat{{X}}_{j}+\Delta \hat{{X}}_{j}\Delta \hat{{X}}_{i}\rangle /2  \label{cov}
\end{equation}
with $\Delta \hat{{X}}_{j}=\hat{{X}}_{j}-\langle \hat{{X}}_{j}\rangle $, $\hat{{x}}_{j}=(\hat{a}_{j}+\hat{a}_{j}^\dag)/\sqrt{2}$, and $\hat{{p}}_{j}=(\hat{a}_{j}-\hat{a}_{j}^\dag)/\sqrt{2}i$. The commutation relations $[\hat{{X}}_{i},\hat{{X}}_{j}]=iU_{ij}$ with $\mathbf{U}=\left(\begin{array}{cc}\mathbf{J} & 0 \\0 & \mathbf{J}\end{array}\right) $ and $\mathbf{J}=\left(\begin{array}{cc}0 & 1 \\-1 & 0 \end{array}\right) $ defines the symplectic structure of the system, which is further characterized by the symplectic eigenvalues $\mathbf{\nu }=(\nu _{1},\nu _{2})$ of the matrix $i\mathbf{U\cdot {V}}$. The Heisenberg's uncertainty principle exerts a constraint on $\nu _{i}$ such that $\nu_{i}\geqslant 1/2$. Thus the Peres-Horodecki criterion \cite{Peres1996, Horodecki1996} is rephrased as that the state is separable whenever the uncertainty principle is still obeyed by the covariance matrix under the partial transposition \cite{Simon2000}, which connects to $\mathbf{V}$ as $\tilde{\mathbf{V}}=\mathbf{\Lambda} \cdot \mathbf{V}\cdot \mathbf{\Lambda }$ with $\mathbf{\Lambda }=diag(1,1,1,-1)$. If the state is entangled, then $\tilde{\mathbf{V}}$ would violate the uncertainty principle and $\tilde{\nu}_{i}$ would be smaller than $1/2$. The logarithmic negativity just quantifies this violation as \cite{Vidal2002}
\begin{equation}
E_{N}=\max \{0,-\log _{2}[2\min (\tilde{\nu}_{1},\tilde{\nu}_{2})]\}.\label{measu}
\end{equation}

In general, the analytical relations among the elements of the obtained $\mathbf{ V}(t)$ is hard to evaluate. Under the condition that two mechanical mirrors are identical, we can prove analytically $\mathbf{{V}}(t)=\left(
\begin{array}{cc}
\mathbf{{A}} & \mathbf{{C}} \\
\mathbf{{C}}^{T} & \mathbf{{A}}%
\end{array}%
\right) $, where $V_{12}(t)=V_{21}(t)$ and $
{\mathbf{C}} =\left(
\begin{array}{cc}
V_{13}(t) & -V_{12}(t) \\
-V_{12}(t) & {V}_{24}(t)%
\end{array}%
\right)$
with $V_{11}(t)+V_{13}(t)=V_{22}(t)+ V_{24}(t)=\bar{n}_{0}+1/2$  (see Appendix \ref{app-covariance}). Here we have assumed that the oscillators are initially in thermal equilibrium with their respective reservoirs. Thus there are only three independent
variables in $\mathbf{V}(t)$. Defining $\mathbf{V}^{(3)}(t)=\left(
\begin{array}{ccc}
V_{11}(t), & V_{22}(t), & V_{12}(t)%
\end{array}%
\right) ^{T}$, we can calculate from Eq. (\ref{speredms})
\begin{equation}
\dot{\mathbf{V}}^{(3)}(t)=\mathbf{M}^{(3)}\mathbf{\cdot V}^{(3)}(t)+\mathbf{B}^{(3)}(t)  \label{evo7}
\end{equation}%
with
\begin{eqnarray}
&&\mathbf{M}^{(3)}=  \left(
\begin{array}{ccc}
-2\gamma _{0} & 0 & 2\omega _{0} \\
0 &2( 2\zeta_-^r-\gamma _{0}) & 2(2\zeta_-^i-\omega _{0})
\\
2\zeta_-^i-\omega _{0} & \omega _{0} & 2(\zeta_-^r-\gamma _{0})%
\end{array}%
\right) ,  \\
&&\mathbf{B}^{(3)}(t)=\left(
\begin{array}{ccc}
\phi , & \phi +2\xi^r-\frac{\phi \zeta^r _{-}}{\gamma _{0}},
&\xi^i- \frac{\phi \zeta^i _{-}}{2\gamma _{0}}%
\end{array}%
\right) ^{T},  \label{evo9}
\end{eqnarray}%
where $\zeta^r _{-}+i\zeta_-^i=\frac{2\eta _{0}^{2}|\alpha |^{2}}{\kappa
-i(\Delta+\omega _{0})}-\frac{2\eta _{0}^{2}|\alpha |^{2}}{\kappa
+i(\Delta-\omega _{0})}$, $\xi^r+i\xi^i =\eta _{0}^{2}(\xi _{1}^{-}+\xi
_{1}^{+\ast })$, and $\phi =\gamma _{0}(2\bar{n}_{0}+1)$. The initial condition is $\mathbf{V}^{(3)}(0)=\left(
\begin{array}{ccc}
\bar{n}_{0}+{1/2}, & \bar{n}_{0}+{1/2}, & 0%
\end{array}%
\right) ^{T}$.

Except for solving Eq. \eqref{evo7} numerically to evaluate the entanglement, we can also obtain an analytical form of $E_N$. For this purpose, we need to convert $\mathbf{V}(t)$ into a normal form $\bar{\mathbf{V}}(t)=\mathbf{U}(t)\mathbf{\cdot V}(t)\mathbf{\cdot U}^{T}(t)$ by a local unitary transformation $
\mathbf{U}(t)=diag (e^{i\theta\sigma_y/2},~e^{i\theta\sigma_y/2})$ with $\sigma_y$ the Pauli matrix. Such transformation leaves the entanglement unchanged. The achieved $\bar{\mathbf{V}}(t)$ corresponds to the covariance matrix defined in the rotated quadrature vector $\mathbf{\hat{\bar{X}}}\equiv(\hat{\bar{x}}_{1},\hat{\bar{p}}_{1},\hat{\bar{x}}_{2},\hat{\bar{p}}_{2})$ with $\hat{\bar{x}}_j=[\hat{a}_je^{-i\theta/2}+\text{h.c.}]/\sqrt{2}$ and $\hat{\bar{p}}_j=[\hat{a}_je^{-i\theta/2}-\text{h.c.}]/\sqrt{2}i$. Choosing $\theta=\arg [\langle \hat{a}_{1}\hat{a}_{1}\rangle (t)]$, we get a normal form $\bar{\mathbf{V}}(t)=\left(
\begin{array}{cc}
\bar{\mathbf{A}} & \bar{\mathbf{C}}\\
\bar{\mathbf{C}}^{T} & \bar{\mathbf{A}}%
\end{array}%
\right) $, where
\begin{eqnarray}
\bar{\mathbf{A}} =diag\left(\bar{V}_{11}(t),\bar{V}_{22}(t)\right),
\bar{\mathbf{C}} =diag\left(\bar{V}_{13}(t),\bar{V}_{24}(t)\right)  ~~
\end{eqnarray}%
under $\bar{V}_{11}(t)+\bar{V}_{13}(t)=\bar{V}_{22}(t)+\bar{V}_{24}(t)=\bar{n}_{0}+{1/2}$. Then the two symplectic eigenvalues of $\tilde{\mathbf{V}}(t)$ are
\begin{equation}
\tilde{\nu}_{1,2}=\{[\bar{V}_{11}(t)\pm\bar{V}_{13}(t)][\bar{V}_{22}(t)\mp\bar{V}_{24}(t)]\}^{{1\over2}}.\label{nu12}
\end{equation}
In terms of the center-of-mass and relative motion quadrature operators $\hat{\bar{Q}}_{\pm }=(\hat{\bar{x}}_{1}\pm \hat{\bar{x}}_{2})/\sqrt{2}$ and $\hat{\bar{P}}_{\pm }=(\hat{\bar{p}}_{1}\pm \hat{\bar{p}}_{2})/\sqrt{2}$, we have $\bar{V}_{11}(t)\pm \bar{V}_{13}(t)=\delta \bar{Q}_{\pm }^{2}(t)$ and $\bar{V}_{22}(t)\pm \bar{V}_{24}(t)=\delta \bar{P}_{\pm }^{2}(t)$. Thus Eq. (\ref{nu12}) changes into
\begin{eqnarray}
\tilde{\nu}_1&=&[(\bar{n}_0+1/2)\delta \bar{P}_{-}^{2}(t)]^{{1\over2}},\\ \tilde{\nu}_2&=&[(\bar{n}_0+1/ 2)\delta \bar{Q}_{-}^{2}(t)]^{{1\over2}}.~~~\label{smpev}
\end{eqnarray}
The identities $\delta \bar{P}_{-}^{2}(t)=2\bar{V}_{22}(t)-\delta\bar{P}_+^2$ and $\delta \bar{Q}_{-}^{2}(t)=2\bar{V}_{11}(t)-\delta\bar{Q}_+^2$ lead to $\delta \bar{P}_{-}^{2}(t)-\delta \bar{Q}_{-}^{2}(t)=2[\bar{V}_{22}(t)-\bar{V}_{11}(t)]$. From the definition of $\bar{\mathbf{V}}(t)$, we have $\bar{V}_{11}(t)-\bar{V}_{22}(t)=2\text{Re}[\langle\hat{a}_{1}\hat{a}_{1}\rangle (t)e^{-i\theta }]$. Remembering $\theta=\arg [\langle \hat{a}_{1}\hat{a}_{1}\rangle (t)]$, we readily obtain $\bar{V}_{11}(t)-\bar{V}_{22}(t)=2|\langle \hat{a}_{1}\hat{a}_{1}\rangle(t)|\geq 0$. Therefore, we always have $\tilde{\nu}_1<\tilde{\nu}_2$. Thus the non-zero entanglement in Eq. (\ref{measu}) is achievable only when
\begin{equation}
\tilde{\nu}_1<1/ 2\Rightarrow \delta \bar{P}_{-}^{2}(t)< 1/[2(2\bar{n}_{0}+1)]  .\label{entcri}
\end{equation}
Equation (\ref{entcri}) indicates that the entanglement between the oscillators is only determined by the variances of their relative momentum operator $\hat{\bar{P}}_-(t)$, which is only sensitive to the environmental temperature.

From the commutation relation $[\hat{\bar Q}_-,\hat{\bar P}_-]=i$, we have their uncertainty relation $\delta \bar{Q}_{-}^{2}\delta \bar{P}_{-}^{2}\geq 1/4$. Therefore, the squeezing property is present when either $\delta \bar{Q}_{-}^{2}<1/2$ or $\delta \bar{P}_{-}^{2}<1/2$. Thus we can conclude from the analytical result in Eq. (\ref{entcri}) that it is the squeezing in the relative momentum $\hat{\bar P}_-$ that is responsible for the entanglement generation of the mechanical oscillators. At zero temperature, i.e. $\bar{n}_{0}=0$, the entanglement between the two oscillators can be established whenever the squeezing shows up in $\hat{\bar{P}}_-$; while at finite temperature, the generation of entanglement needs stronger squeezing in $\hat{\bar{P}}_-$.

\subsection{Numerical verification}
\begin{figure}[tbp]
\includegraphics[width=\columnwidth]{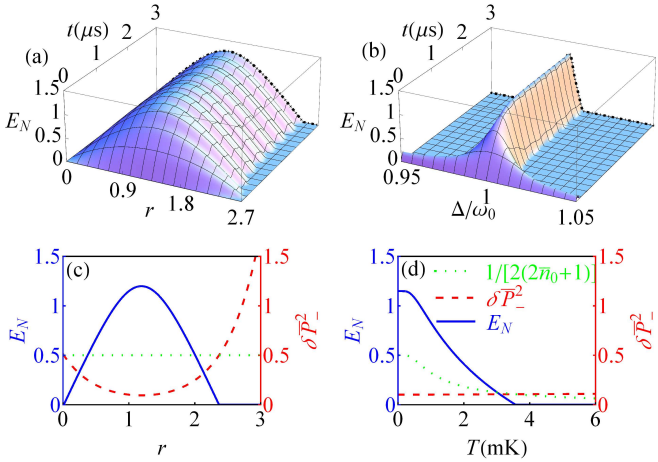}
\caption{(Color online) Entanglement dynamics in different squeezing parameter $r$ when $\Delta/\omega _{0}=1$ (a) and in different $\Delta$ when $r=1$ (b), where the black dot-dashed lines show the results from the steady-state solution. The steady-state entanglement $E_{N}$ (blue solid line) and the squared variance $\delta \bar{P}_{-}^{2}$ (red dashed line) in different squeezing parameter $r$ when $T=0$ K (c) and in different temperature $T$ when $r=1$ (d), where the frequency detuning is chosen as $\Delta=\omega _{0}$ and the green dotted lines show the value $1/[2(2\bar{n}_0+1)]$. $P=4\mu$W and other parameters are given in the main text.} \label{Fig2}
\end{figure}
To verify our analytical expectation, we numerically calculate the covariance matrix from Eq. (\ref{evo7}) and evaluate its entanglement. The parameters are chosen as follows: The cavity-field frequency is $\omega _\text{c}=2\pi \times 6.98\times 10^{9}$ Hz, its damping rate is $\kappa =2\pi \times 6.2\times 10^{6}$ Hz, the frequency of the mechanical oscillators is $\omega _{0}=2\pi \times 32.1\times 10^{6}$ Hz, their damping rate is $\gamma _{0}=15\times 10^{-5}\kappa $, and the coupling strength is $\eta _{0}=2\pi \times 39$ Hz. These parameters are accessible with the recent experiment \cite{Massel2012}. Figure \ref{Fig2}(a) shows the entanglement evolution in different squeezing parameter $r$. Obviously, when $r=0$, which corresponds to the broadband driving field being a vacuum reservoir, no entanglement can be stimulated. However, with the switching on the squeezing of the broadband driving field, a notable entanglement can be induced asymptotically in a wide range of $r$. This demonstrates well that the squeezing in the broadband driving field to the cavity field is responsible for the entanglement generation of the two mechanical oscillators. Figure \ref{Fig2}(b) shows the entanglement dynamics in different detuning $\Delta$ of the broadband driving field to the coherent driving field. We can see that the best performance of the squeezing field on generating entanglement is achieved when the central frequency detuning $\Delta$ of the squeezed field matches with the frequency of the oscillators. We also can observe that the entanglement does not tend to a constant value in the long-time limit, but a lossless oscillation with tiny amplitude. This oscillation in frequency $\Delta$ originates from the time-dependent dissipator introduced by the broadband squeezed field, as shown by $\hat{\mathcal{L}}_\text{c}$ in Eq. (\ref{mastsqu}) and in Eq. (\ref{evo7}) with the explicit form of $\mathbf{B}^{(3)}(t)$ in Eq. (\ref{appb3}). In the following, we take $t={Z\pi/2\Delta}$, where $Z$ is an arbitrarily large integer to ensure the system arriving at its steady state, to study the steady-state properties of the system. Figure \ref{Fig2}(c) plots the calculated steady-state quantum entanglement $E_N$ and the squared variance $\delta \bar{P}_{-}^{2}(\infty)=2\bar{V}_{22}(\infty)-(\bar{n}_{0}+1/2)$ at zero temperature (i.e. $\bar{n}_0=0$) in different squeezing parameter $r$. It indicates clearly that the region where the stable entanglement is formed matches well with the region where $\delta\bar{P}_-^2<1/2$. At finite temperature, Fig. \ref{Fig2}(d) shows that, whenever $\delta\bar{P}_-^2<1/[2(2\bar{n}_0+1)]$ is satisfied, a non-zero entanglement can be established. All these results prove the validity of the entanglement criterion \eqref{entcri}. It is also impressive to find from Fig. \ref{Fig2}(d) that the dramatic entanglement can be generated even when the environmental temperature is in the order of magnitude of mK, where one generally believes that the dominate thermal fluctuation would destroy quantum effects. Here it is remarkable that the quantum correlation is triggered alive even in this temperature region.

Both of our analytical and numerical results reveal that the cavity field acts as a quantum bus to transfer the squeezing character from the broadband driving field to the relative motion degree of freedom of the two mechanical oscillators such that a stable nonclassical correlation can be established between them. It suggests an interesting way to generate stable entanglement between the mechanical mirrors by engineering the squeezing property of the reservoir felt by the cavity field. The physical reason why the relative-motion operator instead of center-of-mass one is squeezed by the cavity field can be understood in the following way. In our system configuration, $\eta_1=-\eta_2\equiv \eta_0$, under which Eq. (\ref{Hamini}) is converted to
\begin{equation}
\hat{H}=\Delta_\text{c}\hat{c}^\dag\hat{c}+\omega_0\sum_{j=\pm}\hat{a}^\dag_j\hat{a}_j+\sqrt{2}\eta_0\hat{Q}_-\hat{c}^\dag\hat{c}+\Omega(\hat{c}^\dag+\hat{c}).
\end{equation}
It indicates that only the relative coordinate of the oscillators feels the existence of the cavity field, while the center-of-mass coordinate as a dark mode is immune to the cavity field. Thus the squeezing character of the reservoir is transferred to relative quadrature operators via its interaction with the cavity field (see Fig. \ref{FigS2}). This again justifies the validity of our analytic result in (\ref{entcri}). The mediation role of the common cavity field here relates to the scheme in Ref. \cite{Li2015}, where an effective squeezing of the two mechanical modes is achieved by applying two coherent driving fields to the cavity, while in ours it is transferred from the broadband squeezed reservoir via the cavity field.

\begin{figure}[tbp]
\centering
\includegraphics[width=\columnwidth]{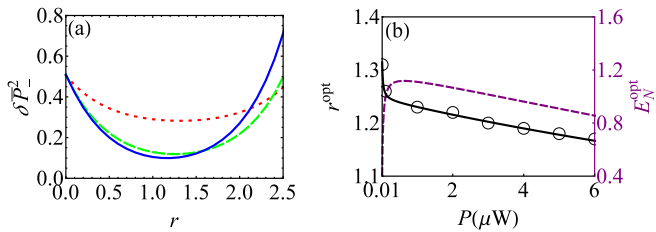}
\caption{(Color online) (a) The squared variance $\delta \bar{P}_{-}^{2}$ as a function of $r$ when $P=0.01$ (red dotted line), $0.1$ (green dashed line), and $2\mu$W (blue solid line). (b) The optimal squeezing $r^{\text{opt}}$ (black solid line) to obtain the smallest $\delta \bar{P}^2_-(\infty)$ and the corresponding steady-state entanglement $E_{N}^{\text{opt}}$ (purple dashed line) as a function of the driving power $P$. The black circles are calculated from numerical solution and the black solid line is calculated from Eq. (\ref{opt}). Parameters as $T=0$ K, $\Delta/\omega _{0}=1$ have been used and the others are shown in the main text.}
\label{Fig3}
\end{figure}

Since the entanglement originates from the squeezing in $\hat{\bar P}_-$, a largest entanglement is achievable when $\delta{\bar P}^2_-(\infty)$ has a smallest value. To get the best performance of our scheme on entanglement generation, we now explore the optimal condition of the system parameters on entanglement generation. Figure \ref{Fig3}(a) shows the calculated $\delta \bar{P}_{-}^{2}(\infty)$ as a function of $r$ in different driving power $P$, which indicates that the optimal value of $r$ to get the smallest $\delta{\bar P}^2_-(\infty)$ is dependent on the driving power $P$. Via $d\delta^2\bar{P}_-^2(\infty)/dr=2d\bar{V}_{22}(\infty)/dr=0$, we have the optimal $r$ to make $\delta^2\bar{P}_-^2$ smallest (see Appendix \ref{solucm})
\begin{equation}
r^{\text{opt}}=\frac{1}{2}\text{arctanh} \{\frac{\mathbf{\Theta }\cdot \text{Re}%
[(2i\Delta\mathbf{I}-\mathbf{M}^{(3)})^{-1}\cdot \mathbf{B}%
_{2}e^{2i\Delta t}]}{\mathbf{\Theta }\cdot [\mathbf{M}^{(3)}]^{-1}\cdot \mathbf{B}_{1}}\}.  \label{opt}
\end{equation}
Figure \ref{Fig3}(b) shows $r^{\text{opt}}$ obtained from numerical calculation and from Eq. \eqref{opt} and the corresponding optimal steady-state entanglement as a function of the pumping power $P$. As we can see, with the increase of $P$, $r^{\text{opt}}$ decreases and the optimal entanglement generated increases and saturates at a moderate pumping power. This indicates that a moderate pumping suffices the generation of a maximal entanglement between mechanical oscillators.

\begin{figure}[tbp]
\includegraphics[width=\columnwidth]{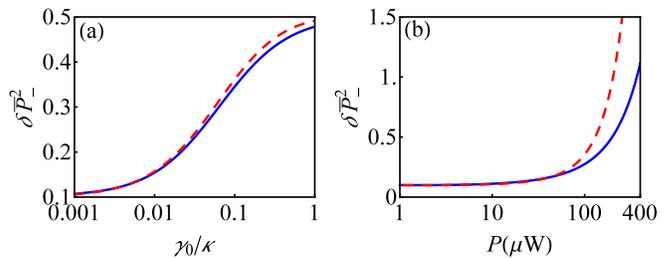}
\caption{(Color online) $\delta \bar{P}_{-}^{2}(\infty)$ evaluated from Eq. \eqref{mastini} with (blue solid line) and from Eq. \eqref{speredms} without the adiabatic elimination (red dashed line) when $P=4\mu$W in (a) and $\gamma _{0}/
\kappa =1$ in (b). Parameters $T=0$K, $\Delta/\omega _{0}=1$, and $r=1$ have been used and the others are shown in the main text.}
\label{Fig4}
\end{figure}
Before closing our discussion, we verify the validity of the adiabatic elimination used in the derivation of Eq. \eqref{speredms}. Figure \ref{Fig4} plots $\delta \bar{P}_{-}^{2}$ calculated from Eqs. \eqref{mastini} and \eqref{speredms}, which corresponds to the cases without and with the adiabatic elimination, respectively. It shows that the adiabatic elimination works well in a wide parameter regime, e.g., at large cavity decay rate and small and moderate pumping power. This result proves the validity of our above calculation, where the parameters used are within the permissible scope of the adiabatic elimination.

\section{Conclusions}\label{con}

In summary, we have proposed a scheme to generate stable entanglement between the two mechanically oscillating mirrors of a cavity by engineering the squeezing character of the reservoir felt by the cavity. Via adiabatically eliminating the degree of freedom of the cavity field, a reduced master equation satisfied by the two mechanical oscillators is derived microscopically. From this master equation, we have analytically found that the generated entanglement of the two mirrors originates from the squeezing of the relative momentum of the two mirrors. Our result reveals that the cavity field acts as a quantum bus to transfer the squeezing character of its reservoir to the relative momentum of the two mechanical oscillators such that a stable entanglement is established in their steady state. The numerical verification indicates that our proposal is realizable with the present experimental technique of cavity optomechanics.

\section*{ACKNOWLEDGEMENTS}
This work is supported by the Fundamental Research Funds for the Central Universities, by the Specialized Research Fund for the Doctoral Program of Higher Education, by the Program
for New Century Excellent Talents in University, and by the National Natural Science Foundation of China (Grants No. 11175072, No. 11474139, No. 11422437, and No. 11574353).

\appendix

\section{The derivation of the reduced master equation}\label{derimse}
Working in the interaction picture and further introducing a squeezing transformation, we can recast Eq. (\ref{mastini}) into \cite{Cirac1992,Swain1998}
\begin{equation}
\dot{\bar{W}}(t)=-i[\hat{\bar{H}}_{\text{I}}(t),\bar{W}(t)]+\mathcal{\hat{L}}_{\text{m}}\bar{W}(t)+\mathcal{\hat{L}}_{\text{vac}}\bar{W}(t),\label{mastsqu}
\end{equation}
where $\bar{W}(t)=\hat{S}\exp (i\hat{H}_{0}t)W(t)\exp (-i\hat{H}_{0}t)\hat{S}^{\dag }$ with $\hat{S}=\exp [r(\hat{c}^{2}-\hat{c}^{\dag 2})/2]$, $\mathcal{\hat{L}}_{\text{vac}}\cdot =\kappa \check{\mathcal{D}}_{\hat{c},\hat{c}^{\dag }}\cdot $, $\hat{\bar{H}}_{\text{I}}(t)=\hat{A}^{\dag }(t)\hat{c}+\text{h.c.}$ with $\hat{A}(t)=\sum_{j}\tilde{\eta}_{j}(t)[\hat{a}
_{j}e^{-i\omega _{j}t}+\text{h.c.}]$ and $\tilde{\eta}_{j}(t)=\eta_{j}(\alpha ^{\ast }\sqrt{N}e^{-i\Delta t}+\alpha \sqrt{N+1}e^{i\Delta t})$. Note that $\Delta_{\text{c}}=\Delta _{\text{s}}\equiv \Delta$ has been used in the above transformation.

Governed by the dissipator $\mathcal{\hat{L}}_{\text{vac}}$, the cavity field rapidly approaches the steady state $(|0\rangle\langle 0|)_\text{c}$ in the large damping limit ($\kappa \gg \gamma _{j}$). It means that $\bar{W}(t)$ approximately factorizes as $\bar{W}(t)\backsimeq \text{Tr}_\text{c}[\bar{W}(t)]\otimes (|0\rangle\langle 0|)_\text{c}$. Seeing the cavity field as ``a reservoir", we can adiabatically eliminate its degree of freedom and obtain a reduced master equation satisfied by the two oscillators. Explicitly, in the dissipation picture $\tilde{W}(t)=\exp [-\mathcal{\hat{L}}_{\text{vac}}t]\bar{W}(t)$, Eq. (\ref{mastsqu}) can be recast into $\dot{\tilde{W}}(t)=[\hat{\tilde{\mathcal{L}}}_{\text{I}}(t)+\hat{\mathcal{L}}_{\text{m}}]\tilde{W}(t)$, where $\hat{\tilde{\mathcal{L}}}_{\text{I}}(t)\cdot =-i\exp (-\hat{\mathcal{L}}_{\text{vac}}t)[\hat{\bar{H}}_{\text{I}}(t),\cdot ]\exp (\hat{\mathcal{L}}_{\text{vac}}t)$. Under the Born approximation $\tilde{W}(t)=\tilde{\rho}(t)\otimes(|0\rangle \langle 0|)_{\text{c}}$ and the Markovian approximation, we obtain
\begin{equation}
\dot{\tilde{\rho}}(t)=\hat{\mathcal{L}}_{\text{m}}\tilde{\rho}(t)+\text{Tr}_{\text{c}}\int_{0}^{\infty }d\tau \hat{\tilde{\mathcal{L}}}_{\text{I}}(t)\hat{\tilde{\mathcal{L}}}_{\text{I}}(t-\tau )\tilde{\rho}(t)(|0\rangle \langle0|)_{\text{c}},  \label{mastsec}
\end{equation}
where $\tilde{\rho}(t)=\text{Tr}_{\text{c}}[\tilde{W}(t)]$ and $\hat{\tilde{\mathcal{L}}}_{\text{I}}(t)\cdot =-i[\mathcal{\hat{A}}_{+}(t)%
\mathcal{\hat{C}}_{-}(t)+\mathcal{\hat{A}}_{-}(t)\mathcal{\hat{C}}_{+}(t)-%
\text{h.c.}]\cdot$
with $\mathcal{\hat{A}}_{+}(t)\cdot =\hat{A}^{\dag }(t)\cdot $, $\mathcal{%
\hat{A}}_{-}(t)\cdot =\hat{A}(t)\cdot $, $\mathcal{\hat{C}}_{-}(t)\cdot
=\exp (-\hat{\mathcal{L}}_{\text{vac}}t)(\hat{c}\cdot )\exp (\hat{\mathcal{L}%
}_{\text{vac}}t)$, and $\mathcal{\hat{C}}_{+}(t)\cdot =\exp (-\hat{\mathcal{L%
}}_{\text{vac}}t)(\hat{c}^{\dag }\cdot )\exp (\hat{\mathcal{L}}_{\text{vac}%
}t)$. Making a time derivative to $\mathcal{\hat{C}}_{-}(t)$, we have $d\mathcal{\hat{C}}_{-}(t)/dt=-e^{-\hat{\mathcal{L}}_{\text{vac}}t}[%
\hat{\mathcal{L}}_{\text{vac}},\hat{c}\cdot ]e^{\hat{\mathcal{L}}_{\text{vac}%
}t}$. One can easily check $[\hat{\mathcal{L}}_{\text{vac}},\hat{c}\cdot
]=\kappa \hat{c}\cdot $. Thus
\begin{equation}
\mathcal{\hat{C}}_{-}(t)=e^{-\kappa t}(\hat{c}\cdot ),  \label{cmt1}
\end{equation}%
which also induces
\begin{equation}
\mathcal{\hat{C}}_{-}^{\dag }(t)=e^{-\kappa t}(\cdot \hat{c}^{\dag }).
\label{cmt}
\end{equation}%
With the similar manner, we have $d\mathcal{\hat{C}}_{+}(t)/dt=-e^{-\hat{\mathcal{L}}_{\text{vac}}t}[%
\hat{\mathcal{L}}_{\text{vac}},\hat{c}^{\dag }\cdot ]e^{\hat{\mathcal{L}}_{%
\text{vac}}t}$.
From the commutation relation $[\hat{\mathcal{L}}_{\text{vac}},\hat{c}^{\dag
}\cdot ]=2\kappa \cdot \hat{c}^{\dag }-\kappa \hat{c}^{\dag }\cdot $, it can be recast into
\begin{equation}
d\mathcal{\hat{C}}_{+}(t)/dt=-2\kappa \mathcal{\hat{C}}_{-}^{\dag
}(t)+\kappa \mathcal{\hat{C}}_{+}(t).
\end{equation}%
In the form of Eq. (\ref{cmt}), we can obtain
\begin{equation}
\mathcal{\hat{C}}_{+}(t)=e^{\kappa t}(\hat{c}^{\dag }\cdot )+(e^{-\kappa
t}-e^{\kappa t})(\cdot \hat{c}^{\dag }),  \label{cpt1}
\end{equation}%
which also results in
\begin{equation}
\mathcal{\hat{C}}_{+}^{\dag }(t)=e^{\kappa t}(\cdot \hat{c})+(e^{-\kappa
t}-e^{\kappa t})(\hat{c}\cdot ).  \label{cpt2}
\end{equation}%
From the obtained forms of Eqs. (\ref{cmt1}, \ref{cmt}, \ref{cpt1}, \ref{cpt2}%
), we have the non-zero correlation functions of the cavity field as the
following
\begin{eqnarray}
&&\langle \mathcal{\hat{C}}_{-}(t)\mathcal{\hat{C}}_{+}(t-\tau )\rangle
=\langle \mathcal{\hat{C}}_{+}(t)\mathcal{\hat{C}}_{+}^{\dag }(t-\tau
)\rangle  \notag \\
&=&\langle \mathcal{\hat{C}}_{-}^{\dag }(t)\mathcal{\hat{C}}_{+}^{\dag
}(t-\tau )\rangle =\langle \mathcal{\hat{C}}_{+}^{\dag }(t)\mathcal{\hat{C}}%
_{+}(t-\tau )\rangle =e^{-\kappa \tau }  \label{aplt}
\end{eqnarray}%
where $\langle \cdot \rangle =\text{Tr}_{\text{c}}[\cdot (|0\rangle \langle
0|)_{\text{c}}]$.

Substituting Eq. (\ref{aplt}) into Eq. (\ref{mastsec}), we can obtain
\begin{eqnarray}
&&\text{Tr}_{\text{c}}\int_{0}^{\infty }d\tau \hat{\tilde{\mathcal{L}}}_{%
\text{I}}(t)\hat{\tilde{\mathcal{L}}}_{\text{I}}(t-\tau )\tilde{\rho}%
(t)(|0\rangle \langle 0|)_{\text{c}}=\int_{0}^{\infty }d\tau e^{-\kappa \tau
}  \notag \\
&\times &[\hat{A}(t)\tilde{\rho}(t)\hat{A}^{\dag }(t-\tau )-\hat{A}^{\dag
}(t)\hat{A}(t-\tau )\tilde{\rho}(t)+\text{h.c.}]  \label{dercor}
\end{eqnarray}%
Remembering the form of $\hat{A}(t)$ and returning back to the Schr\"{o}%
dinger picture, we have%
\begin{eqnarray}
&&\int_{0}^{\infty }e^{-\kappa \tau }e^{-i\hat{H}_{0}t}\hat{A}(t)\tilde{\rho}(t)%
\hat{A}^{\dag }(t-\tau )e^{i\hat{H}_{0}t}d\tau  \notag \\
&=&\sum_{j,k}\eta _{j}\eta _{k}[(\hat{a}_{j}+\hat{a}_{j}^{\dag })\rho
(t)(\xi _{k}^{+}\hat{a}_{k}^{\dag }+\xi _{k}^{-}\hat{a}_{k})\},
\label{apdxder}
\end{eqnarray}%
with $\xi _{k}^{\pm }=\frac{\digamma }{\kappa \pm i(\omega _{k}+\Delta )}+\frac{|\alpha |^{2}+\digamma ^{\ast }}{\kappa \pm i(\omega
_{k}-\Delta)}$ and $\digamma =N|\alpha |^{2}+M\alpha
^{2}e^{2i\Delta t}$. In obtaining Eq. (\ref{apdxder}), the
integral identity $\int_{0}^{\infty }\exp [-\kappa \tau -i\omega \tau ]d\tau
=1/(\kappa +i\omega )$ has been used. The other terms in Eq. (\ref{dercor})
can be calculated in the similar manner. Then the final form the reduced
master equation (\ref{remast}) can be obtained.

\section{Dynamical evolution of the system}\label{app-dynamics}
\begin{figure}[tbp]
\centering
\includegraphics[width=0.8\columnwidth]{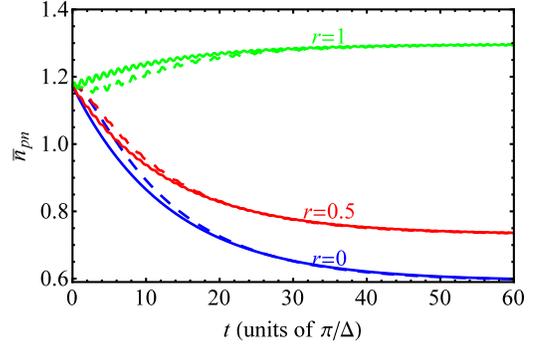}
\caption{(Color online) Dynamical evolution of the mean phonon number calculated with (solid lines) and without (dashed lines) the adiabatic elimination in different values of $r$. Parameters $T=2.5$mK, $\Delta/\omega_0=1$, and $\gamma_0/\kappa=1.5\times10^{-3}$ have been used and the others are shown in the main text.}
\label{FigS1}
\end{figure}
\begin{figure}[tbp]
\centering
\includegraphics[width=\columnwidth]{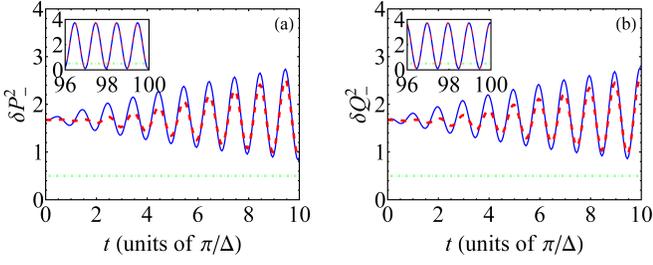}
\caption{(Color online) Fluctuation dynamics of $\delta P_-^2$ in (a) and $\delta Q_-^2$ in (b) calculated with (blue solid lines) and without (red dashed lines) the adiabatic elimination. Insets are results in the long-time regime. Parameter $r=1$ has been used and the others are the same in Fig. \ref{FigS1}.}
\label{FigS2}
\end{figure}
The dynamics of the system can be studied readily by the derived reduced master equation \eqref{speredms}, from which we can see clearly that the center-of-mass motion is decoupled to the cavity mode, while the relative motion is strongly affected by the optomechanical coupling. For the relative motion, not only the thermal dissipation, but also the squeezing-like dissipation is triggered. To verify the validity of Eq. \eqref{speredms}, we plot in Fig. \ref{FigS1} the evolution of the mean phonon number $\bar{n}_{pn}=\text{Tr}[ \hat{a}_{j}^{\dag }\hat{a}_{j}\rho(t)] (j=1,2)$ in different values of $r$ obtained from Eq. \eqref{mastini} and from Eq. \eqref{speredms}. We find that the steady-state phonon number increases with the increase of $r$. This is due to the strengthen of the radiation pressure acting on the mirrors, which comes from the increase of the photon number in the cavity by the squeezed-vacuum reservoir. In addition, faint oscillations appear in the long-time limit when $r\neq0$, which originates from the time-dependent dissipator introduced by the broadband squeezed field. Furthermore, we find that the adiabatic elimination performs well during time evolution except for small deviation in the short-time scale. Figure \ref{FigS2} plots the fluctuation dynamics of the relative motion quadrature operators $\hat{P}_{-}=(\hat{a}_{-}-\hat{a}_{-}^{\dag })/\sqrt{2}i$ and $\hat{Q}_{-}=(\hat{a}_{-}+\hat{a}_{-}^{\dag })/\sqrt{2}$. The squeezing is achievable when either $\delta\hat{P}^{2}_{-}(t)<1/2$ or $\delta\hat{Q}^{2}_{-}(t)<1/2$. The results indicate that the quadrature operators can be squeezed in long-time limit, which originates from the transfer of squeezing properties from the reservoir. We also observe that behaviors of the quadrature operators are antiphase with each other and the oscillations keeps in a frequency $2\Delta$.

\section{Covariance matrix of the mechanical oscillators}

\label{app-covariance}

\subsection{The covariance matrix}

Defining a vector $\mathbf{V}^{\text{(F)}}(t)=(V_{11}(t)$,$V_{22}(t)$,$V_{33}(t)$,$V_{44}(t)$ ,$V_{12}(t)$,$V_{13}(t)$,$V_{14}(t),$$V_{23}(t)$,$V_{24}(t)$,$V_{34}(t)$$)^{T}$ and  under the conditions $\ \omega _{1}=\omega _{2}=\omega _{0}$, $\gamma_{1}=\gamma _{2}\equiv\gamma _{0}$, $\eta _{1}=-\eta _{2}\equiv\eta _{0}$, and $\bar{n}_{1}=\bar{n}_{2}\equiv\bar{n}_{0}$, we can derive its time evolution equation from Eq. \eqref{speredms}
\begin{equation}
\mathbf{\dot{V}}^{\text{(F)}}(t)=\mathbf{M}\cdot \mathbf{V}^{\text{(F)}}(t)+\mathbf{B}(t)  \label{evo1}
\end{equation}%
with
\begin{widetext}
\begin{eqnarray*}
\mathbf{M} &=&\left(
\begin{array}{cccccccccc}
-2\gamma _{0} & 0 & 0 & 0 & 2\omega _{0} & 0 & 0 & 0 & 0 & 0 \\
0 & 2(\zeta _{-}^{r}-\gamma _{0}) & 0 & 0 & 2(\zeta _{-}^{i}-\omega _{0}) & 0
& 0 & -2\zeta _{-}^{i} & -2\zeta _{-}^{r} & 0 \\
0 & 0 & -2\gamma _{0} & 0 & 0 & 0 & 0 & 0 & 0 & 2\omega _{0} \\
0 & 0 & 0 & 2(\zeta _{-}^{r}-\gamma _{0}) & 0 & 0 & -2\zeta _{-}^{i} & 0 &
-2\zeta _{-}^{r} & 2(\zeta _{-}^{i}-\omega _{0}) \\
\zeta _{-}^{i}-\omega _{0} & \omega _{0} & 0 & 0 & \zeta _{-}^{r}-2\gamma
_{0} & -\zeta _{-}^{i} & -\zeta _{-}^{r} & 0 & 0 & 0 \\
0 & 0 & 0 & 0 & 0 & -2\gamma _{0} & \omega _{0} & \omega _{0} & 0 & 0 \\
-\zeta _{-}^{i} & 0 & 0 & 0 & -\zeta _{-}^{r} & \zeta _{-}^{i}-\omega _{0} &
\zeta _{-}^{r}-2\gamma _{0} & 0 & \omega _{0} & 0 \\
0 & 0 & -\zeta _{-}^{i} & 0 & 0 & \zeta _{-}^{i}-\omega _{0} & 0 & \zeta
_{-}^{r}-2\gamma _{0} & \omega _{0} & -\zeta _{-}^{r} \\
0 & -\zeta _{-}^{r} & 0 & -\zeta _{-}^{r} & -\zeta _{-}^{i} & 0 & \zeta
_{-}^{i}-\omega _{0} & \beta _{1} & 2(\zeta _{-}^{r}-\gamma _{0}) & -\zeta
_{-}^{i} \\
0 & 0 & \zeta _{-}^{i}-\omega _{0} & \omega _{0} & 0 & -\zeta _{-}^{i} & 0 &
-\zeta _{-}^{r} & 0 & \zeta _{-}^{r}-2\gamma _{0}%
\end{array}%
\right) , \\
 &&\mathbf{B}(t)=\left(
\begin{array}{cccccccccc}
\phi , & \phi +2\xi ^{r}, & \phi , & \phi +2\xi ^{r}, & \xi ^{i}, & 0, &
-\xi ^{i}, & -\xi ^{i}, & -2\xi ^{r}, & \xi ^{i}%
\end{array}%
\right) ^{T},
\end{eqnarray*}
\end{widetext}where $\zeta^r _{-}+i\zeta_-^i=\frac{2\eta _{0}^{2}|\alpha |^{2}}{\kappa
-i(\Delta+\omega _{0})}-\frac{2\eta _{0}^{2}|\alpha |^{2}}{\kappa
+i(\Delta-\omega _{0})}$, $\xi^r+i\xi^i =\eta _{0}^{2}(\xi _{1}^{-}+\xi
_{1}^{+\ast })$, and $\phi =\gamma _{0}(2\bar{n}_{0}+1)$. Considering that the mechanical oscillators are initially in thermal states
with the same temperature as their reservoirs, we have $V_{ii}(0)=\bar{n}%
_{0}+\frac{1}{2}$ and $V_{ij}(0)=0$ for $i\neq j$.

First, according to Eq. (\ref{evo1}), we have
\begin{widetext}
\begin{equation}
{d\over dt}\mathbf{V}^{(4)}(t)\equiv{d\over dt}\left(
\begin{array}{c}
V_{11}(t)-V_{33}(t) \\
V_{22}(t)-V_{44}(t) \\
V_{12}(t)-V_{34}(t) \\
V_{14}(t)-V_{23}(t)%
\end{array}%
\right) =\left(
\begin{array}{cccc}
-2\gamma _{0} & 0 & 2\omega _{0} & 0 \\
0 & 2(\zeta_-^r -\gamma _{0}) & 2(\zeta _-^i-\omega _{0}) & -2%
\zeta_- ^i \\
\zeta _-^i-\omega _{0} & \omega _{0} & \zeta_-^r-2\gamma _{0} & -%
\zeta_-^r \\
-\zeta_-^i & 0 & -\zeta_-^r & \zeta_-^r-2\gamma _{0}%
\end{array}%
\right)\cdot \mathbf{V}^{(4)}(t).\end{equation}
\end{widetext} From the initial condition $\mathbf{V}^{(4)}(0)=(0,0,0,0)^{T}$%
, its dynamical solution can be solved as $\mathbf{V}^{(4)}(t)=(0,0,0,0)^{T}$, which indicates
\begin{eqnarray}
V_{11}(t) &=&V_{33}(t),~V_{22}(t)=V_{44}(t),  \label{re1} \\
V_{12}(t) &=&V_{34}(t),~V_{14}(t)=V_{23}(t).  \label{re11}
\end{eqnarray}

Further defining $\mathbf{V}^{(2)}(t)=(V_{5}(t),V_{6}(t))^{T}$ with $%
V_{5}(t)=V_{12}(t)+V_{14}(t)$ and $%
V_{6}(t)=V_{11}(t)+V_{13}(t)-V_{22}(t)-V_{24}(t)$, and according to Eq. (\ref%
{evo1}), we have $\mathbf{\dot{V}}^{(2)}(t)=\left(
\begin{array}{cc}
-2\gamma _{0} & -\omega _{0} \\
4\omega _{0} & -2\gamma _{0}%
\end{array}%
\right)\cdot \mathbf{V}^{(2)}(t)$. Under the initial condition $\mathbf{V}%
^{(2)}(0)=(0,0)^{T}$, its dynamical solution can be obtained as $\mathbf{V}^{(2)}(t)=(0,0)^{T}$,
which indicates
\begin{eqnarray}
V_{14}(t) &=&-V_{12}(t),  \label{re21} \\
V_{11}(t)+V_{13}(t) &=&V_{22}(t)+V_{24}(t) .  \label{re22}
\end{eqnarray}

In last, defining $V_{7}(t)=V_{11}(t)+V_{13}(t)$, we have
\begin{equation}
\dot{V}_{7}(t)=-2\gamma _{0}V_{7}(t)+\gamma _{0}(2\bar{n}_{0}+1)
\end{equation}%
under the initial condition $V_{7}(0)=\bar{n}_{0}+\frac{1}{2}$. Its solution
reads $V_{7}(t)=\bar{n}_{0}+\frac{1}{2}$, which is time-independent. Thus
\begin{equation}
V_{11}(t)+V_{13}(t)=\bar{n}_{0}+\frac{1}{2}.  \label{re3}
\end{equation}

With the relations (\ref{re1}, \ref{re11}, \ref{re21}, \ref{re22}, \ref{re3}%
), we have the final form of the covariance matrix
\begin{eqnarray}
\mathbf{V}(t) &=&\left(
\begin{array}{cc}
\mathbf{A} & \mathbf{C} \\
\mathbf{C}^{T} & \mathbf{A}%
\end{array}%
\right),\label{ffv}\\
\mathbf{C} &=&\left(
\begin{array}{cc}
\bar{n}_{0}+\frac{1}{2}-V_{11}(t) & -V_{12}(t) \\
-V_{12}(t) & \bar{n}_{0}+\frac{1}{2}-V_{22}(t)%
\end{array}%
\right) .
\end{eqnarray}%

\subsection{The solution of the covariance matrix}\label{solucm}

The time-dependent inhomogeneous term (\ref{evo9}) can be separated into
\begin{eqnarray}
\mathbf{B}^{(3)}(t) &=&\mathbf{B}_{0}+N\mathbf{B}_{1}+M(\mathbf{B%
}_{2}e^{2i\Delta t}+\text{c.c.}),  \notag \\
\mathbf{B}_{0} &=&\left(
\begin{array}{ccc}
\phi , & \phi +\zeta _{+}-\frac{\phi \zeta^r _{-}}{\gamma _{0}}, &
\frac{\gamma _{0}\zeta _{+}^{i}-\phi \zeta _{-}^{i}}{2\gamma _{0}}%
\end{array}%
\right) ^{T},  \notag \\
~~\mathbf{B}_{1} &=&\left(
\begin{array}{ccc}
0, & 2\zeta _{+}^{r}, & \zeta _{+}^{i}
\end{array}%
\right) ^{T}~,  \notag \\
\mathbf{B}_{2} &=&\left(
\begin{array}{ccc}
0, &\bar{\zeta}_{+}, & \frac{i\bar{\zeta}_{-}}{2}
\end{array}%
\right) ^{T}.\label{appb3}
\end{eqnarray}%
where $\zeta^r _{+}+i\zeta^i_+=\frac{2\eta _{0}^{2}|\alpha |^{2}}{\kappa -i(\Delta+\omega _{0})}+\frac{2\eta _{0}^{2}|\alpha |^{2}}{\kappa +i(\Delta-\omega _{0})}$ and $\bar{\zeta}_{\pm }=\frac{2\eta _{0}^{2}\alpha ^{2}}{%
\kappa +i(\Delta+\omega _{0})}\pm \frac{2\eta _{0}^{2}\alpha ^{2}%
}{\kappa +i(\Delta-\omega _{0})}$. With this separation, the dynamical solution of Eq. (\ref{evo7}) can be constructed as
\begin{eqnarray}
\mathbf{V}^{(3)}(t) &=& -\mathbf{Y\cdot D}^{-1}\mathbf{\cdot }(\mathbf{I}-e^{\mathbf{D}t})\mathbf{\cdot Y}^{-1}\mathbf{\cdot (\mathbf{B}_{0}}+N\mathbf{B}_{1})  \notag \\
&&-M[\mathbf{Y\cdot }(\mathbf{D}-2i\Delta\mathbf{I})^{-1}\mathbf{\cdot }(\mathbf{I}e^{2i\Delta t}-e^{\mathbf{D}t})\cdot\mathbf{
Y}\cdot \mathbf {B}_{2}  \notag \\
&&+\text{c.c.}]+ \mathbf{Y\cdot }e^{\mathbf{D}t}\mathbf{\cdot Y}^{-1}\mathbf{\cdot V}^{(3)}(0),  \label{sol-dyn}
\end{eqnarray}%
where $\mathbf{D=Y}^{-1}\cdot \mathbf{M}^{(3)}\cdot \mathbf{Y}$ and $\mathbf{I}$ is a unit matrix. The solution \eqref{sol-dyn} asymptotically approach
\cite{Gu2013}
\begin{eqnarray}
&&\mathbf{V}^{(3)}(\infty ) =-[\mathbf{M}^{(3)}]^{-1}\cdot (\mathbf{B}_{0}+N\mathbf{B}_{1}) \nonumber \\
&&~~~~~~~-M[(\mathbf{M}^{(3)}-2i\Delta\mathbf{I})^{-1}\cdot \mathbf{B}_{2}e^{2i\Delta t}+\text{c.c.}], ~~~~\label{sol-ste}
\end{eqnarray}%
which is a periodically oscillating function in frequency $2\Delta$. From Eq. \eqref{sol-ste}, we can easily check
\begin{equation}
\bar{V}_{22}(\infty)=\mathbf{\Theta}\cdot\mathbf{V}^{(3)}(\infty )
\end{equation}where $\mathbf{\Theta}=\left(
                                       \begin{array}{ccc}
                                         \sin^2(\theta/2) & \cos^2(\theta/2) & -\sin\theta\\
                                       \end{array}
                                     \right)$.

\end{document}